\documentclass[sigconf]{acmart}

\usepackage{framed}
\usepackage{xcolor}
\definecolor{framecolor}{rgb}{0.8,0.2,0.2} 
\renewenvironment{framed}{%
  \MakeFramed{\advance\hsize-\width \FrameRestore}}%
 {\endMakeFramed}

\AtBeginDocument{%
  }

\setcopyright{acmlicensed}
\copyrightyear{2018}
\acmYear{2018}
\acmDOI{XXXXXXX.XXXXXXX}

\acmConference[Conference acronym 'XX]{Make sure to enter the correct
  conference title from your rights confirmation emai}{June 03--05,
  2018}{Woodstock, NY}
\acmISBN{978-1-4503-XXXX-X/18/06}

\copyrightyear{2025}
\acmYear{2025}
\setcopyright{acmlicensed}\acmConference[CHI '25]{CHI Conference on Human Factors in Computing Systems}{April 26-May 1, 2025}{Yokohama, Japan}
\acmBooktitle{CHI Conference on Human Factors in Computing Systems (CHI '25), April 26-May 1, 2025, Yokohama, Japan}
\acmDOI{10.1145/3706598.3713853}
\acmISBN{979-8-4007-1394-1/25/04}

\begin{document}

\title[Breaking Barriers or Building Dependency? Exploring Team-LLM Collaboration in AI-infused Classroom Debate]{Breaking Barriers or Building Dependency? Exploring Team-LLM Collaboration in AI-infused Classroom Debate}

\author{Zihan Zhang}
\authornote{Both authors contributed equally to this research.}
\affiliation{%
  \institution{School of Design, SUSTech}
  \city{Shenzhen}
  \country{China}
}
\email{zhangzihan654@gmail.com}

\author{Black Sun}
\authornotemark[1]
\affiliation{%
 \institution{Aarhus University}
 \city{Aarhus}
 \country{Denmark}
 }
\email{blackthompson770@gmail.com}

\author{Pengcheng An}
\authornote{Corresponding author.}
\affiliation{%
  \institution{School of Design, SUSTech}
  \city{Shenzhen}
  \country{China}
}
\email{anpc@sustech.edu.cn}


\renewcommand{\shortauthors}{Z. Zhang, B. Sun \& P. An}


\begin{abstract}
Classroom debates are a unique form of collaborative learning characterized by fast-paced, high-intensity interactions that foster critical thinking and teamwork. Despite the recognized importance of debates, the role of AI tools, particularly LLM-based systems, in supporting this dynamic learning environment has been under-explored in HCI. This study addresses this opportunity by investigating the integration of LLM-based AI into real-time classroom debates. Over four weeks, 22 students in a Design History course participated in three rounds of debates with support from ChatGPT. The findings reveal how learners prompted the AI to offer insights, collaboratively processed its outputs, and divided labor in team-AI interactions. The study also surfaces key advantages of AI usage—reducing social anxiety, breaking communication barriers, and providing scaffolding for novices—alongside risks, such as information overload and cognitive dependency, which could limit learners' autonomy. We thereby discuss a set of nuanced implications for future HCI exploration.

\end{abstract}

\ccsdesc[500]{Human-centered computing~Empirical studies in HCI}

\ccsdesc[500]{Human-centered computing~Empirical studies in collaborative and social computing}

\keywords{Debate, Collaborative learning, ChatGPT, Human-AI interaction, Classroom}
\begin{teaserfigure}
  \includegraphics[width=\textwidth]{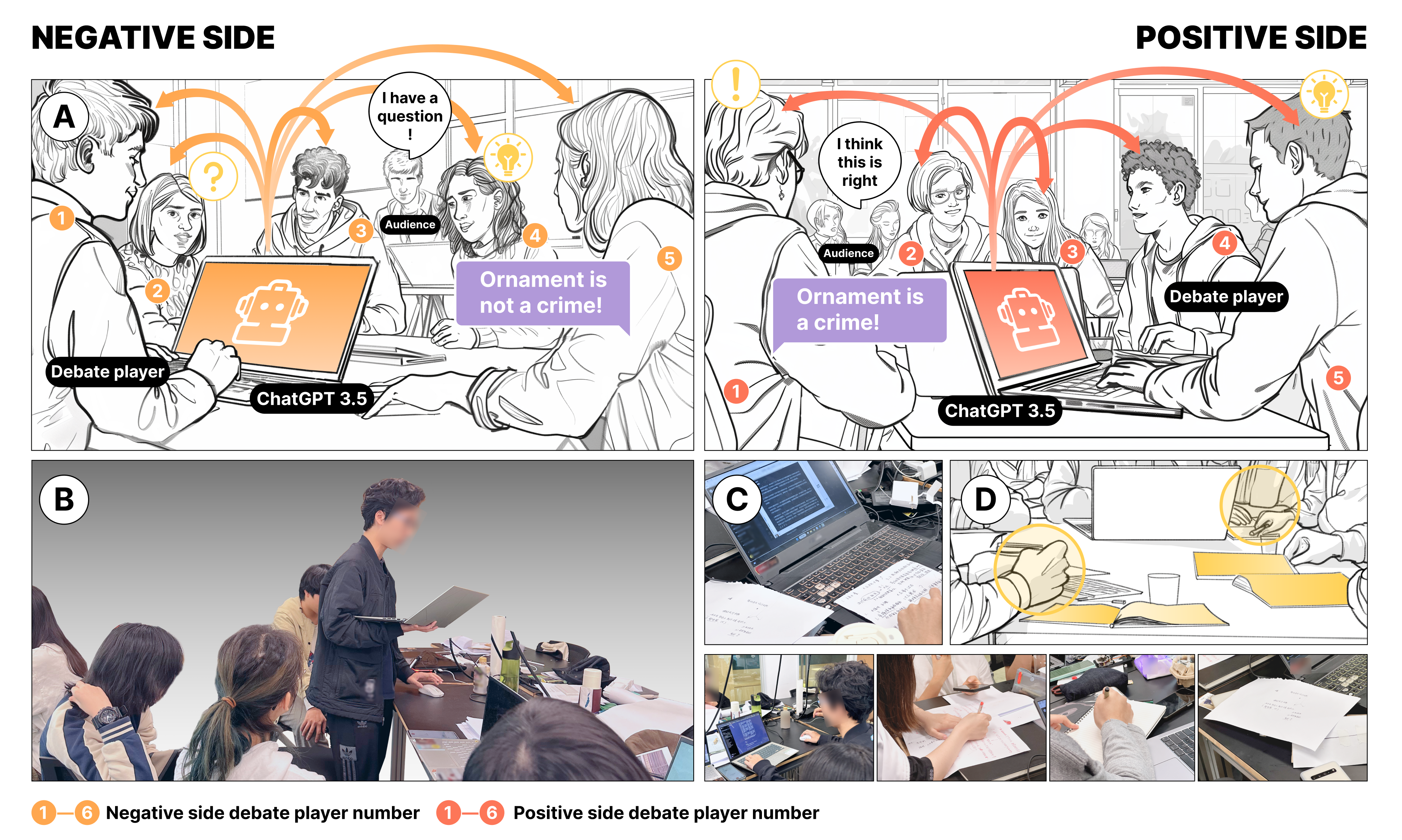}
  \caption{(A) Human-AI Collaboration learning debate (B) Pictures taken at the debate (C) ChatGPT 3.5 assisted debate (D) Various ways to record key points in real-time during the debate}
  \label{fig:teaser}
\end{teaserfigure}


\maketitle

\section{INTRODUCTION}

Researchers are increasingly interested in facilitating human-AI collaboration in complex learning scenarios \cite{amershi2019guidelines, cai2019human}, recognizing AI's potential to transform traditional learning approaches \cite{pataranutaporn2021ai}. Therefore, employing AI in collaborative learning is becoming a significant area of focus. In this domain, a holistic perspective on human-AI partnering is essential, rather than focusing solely on either human or AI capabilities \cite{huang2019human}. Human-AI collaborative learning requires three key characteristics: \textit{"mutual understanding,"} \textit{"mutual benefits,"} and \textit{"mutual growth"} \cite{huang2019human}. While the optimal combination of human expertise and AI capability promises great potential, further exploration is urgently needed to fully comprehend and leverage these interactions \cite{islam2023human}.

Our study aims to gain contextualized insights into human-AI hybrid collaborative learning by focusing specifically on AI-infused classroom debates. Unlike informal debates or general discussions, formal debates are characterized by clear rules and structures, emphasizing the construction of logically rigorous arguments and strategic rebuttals \cite{massey2015introduction}. We chose the classroom debate setting since it represents a high-paced, highly interactive learning activity involving intensive collaboration, communication, and competition within a limited time. This contrasts with the more commonly studied scenarios where human-AI interactions are analytical and relatively less intensive \cite{zheng2024ai}. We believe that examining the dynamics of AI-infused classroom debates can result in unique and meaningful insights that will inform future research on human-AI hybrid collaborative learning across similar contexts, such as classroom peer reviews \cite{mangelsdorf1992peer}, Socratic seminars \cite{tredway1995socratic}, collaborative augmentation \cite{chinn2013learning} and problem-based learning \cite{wood2003problem}.

To do so, we organized 3 rounds of five-to-five classroom debates involving 22 students to understand how learners interact with AI and how these interactions impact their learning process and team collaboration. Each debate centered on a different theme regarding the course Design History, allowing us to observe how participants use AI support in their arguments. We examined learners' emerging ways of interacting with an LLMs-based interface and how these interactions influenced their performance and team dynamics. During the debates, students were required to present their viewpoints and respond to opponents' queries with support from AI-provided examples and perspectives. The entire activity was recorded for analysis. In addition, we conducted in-depth interviews with the 22 participants to explore their views of AI usage during the debates and how this usage influenced their learning behaviors and learning experiences. Analyzing these data provided us with a deeper understanding of the potential benefits and challenges of human-AI collaborative learning in this context. Namely, our investigation encompassed two key research questions:

\textbf{RQ1:} What are the modes of human-AI interaction in debate and collaborative learning?

\textbf{RQ2:} What are AI tools' benefits and potential risks for debate and collaborative learning?

By analyzing these facets, we aim to reveal the impacts of integrating AI into educational debate scenarios. Additionally, the study probed how teams acquired information, formed consensus, and achieved common goals within collaborations and how AI-infused applications served as a mediator to facilitate knowledge sharing. Our findings indicate that AI tools can facilitate the debate and collaborative learning experience by enabling adaptive, tailored responses, offering diverse perspectives, reducing collaborative pressure, and accelerating information retrieval. However, we also observed potential risks, such as cognitive dependency, the undermining of individual creativity, and the challenges of shifting communication expectations. Our research is grounded in an authentic classroom debate environment, offering contextual insights that may guide future research and contribute to designing more effective and considerate human-AI collaborative learning systems for broader educational contexts.

\section{RELATED WORK}

\subsection{Human-AI Hybrid Collaborative and Multi-Technology Support in Education and Learning}\label{sec2.1}

\subsubsection{Traditional AI and Technology in Education}

In the field of education, traditional AI is often applied in areas such as learning analytics \cite{zambrano2024long, dang2024unspoken, nazaretsky2022empowering}, educational data mining \cite{borchers2024you, castro2023understanding}, intelligent tutoring systems (ITS) \cite{weber2024legalwriter, zhou2020improving,tang2023ml4stem, zhang2024mathemyths, borchers2024using, booth2024human}, and learner profiling \cite{mejia2023visualizing}. In these studies, AI is primarily used to analyze vast educational data, generating patterns and insights that help educators and learners reflect on the teaching and learning processes. These systems often provide post-hoc analysis to guide future educational activities or offer evidence-based insights for educational administrators. Some other works are focusing on interactive technologies, such as robots \cite{elmimouni2024navigating}, AR \cite{eisenberg2022increasing, camara2023periodic, zhou2024bee, jin2023shape, caetano2023arlang} and VR \cite{Son2024MakingLE, jin2022will}, cloud-based and client-side applications \cite{underwood2023introducing, chatzidaki2024science, zhang2023experiverse, nunez2023tmbq, ngoon2024classinsight}, dashboards \cite{lee2023multimodal, fernandez2024data, echeverria2024teamslides, buvari2023student, karademir2022designing, gauthier2022adoption}, spatial applications \cite{martinez2022moodoo, fernandez2022classroom}, and documentation tools \cite{tan2024more, sterman2023kaleidoscope}, exploring how these technologies enhance teaching and learning activities. For example, the study by Elmimouni et al. investigated how telepresence robots both enable and limit classroom accessibility \cite{elmimouni2024navigating}. These studies encompass both formal and informal learning, providing rich design cases showing how various media can enhance subject-specific learning across different contexts.

\subsubsection {LLMs in Education and Learning}

In recent years, the rapid development of large language models (LLMs) has driven researchers to actively explore their potential impact on education and learning, gradually integrating them into various aspects of the field. These explorations encompass areas such as programming tutoring \cite{ma2024teach, ross2023programmer, ma2023ai}, children education \cite{lee2023dapie, fu2023self, zhang2024mathemyths, murgia2023chatgpt}, STEM education \cite{chen2024learning, chen2024bidtrainer}, enhancing writing \cite{goldi2024intelligent, mejia2024enhancing, tu2024augmenting, jakesch2023co}, language learning \cite{leong2024putting}, reading assistant \cite{hanel2024towards, xu2023rosita}, classroom interaction \cite{liu2024classmeta, choi2024vivid}, teaching support \cite{tan2024more}, communication learning \cite{shaikh2024rehearsal, liu2023improving} and creative design \cite{cheng2024scientific, harwood2023chai, kocaballi2023conversational}. Based on these works, the roles of LLMs in education can be broadly categorized into three main types. As \textbf{mentors}, LLMs guide learners by providing direct knowledge transmission and skill training, helping them grasp specific concepts and solve problems. As \textbf{collaborators}, LLMs act as interactive partners, engaging with learners to co-create solutions and generate ideas, emphasizing collaboration and creativity. Lastly, as \textbf{assistants}, LLMs support learners and educators by automating tasks, offering personalized recommendations, and tracking learning progress, facilitating teaching and classroom interaction. Additionally, some work has pointed out the opportunities and challenges of LLM in higher education, like efficiency but reducing social interaction \cite{park2024promise}.

\subsubsection{LLMs-infused Multi-learner Collaborative Learning}

However, much of the aforementioned research \cite{ma2024teach, zhang2024mathemyths, goldi2024intelligent, hanel2024towards} focuses on the interaction between LLM systems and individual users, with relatively little attention given to multi-learner collaborative learning. A few examples explore LLMs in group conversation \cite{liu2024peergpt}, group brainstorming \cite{yu2023investigating}, teamwork \cite{zhang2023investigating}, group decision making \cite{chiang2024enhancing}, and classroom simulation \cite{liu2024classmeta}. For example, Liu et al. explored the role of LLM agents in children's collaborative learning, showing that these agents can effectively moderate discussions and foster creative thinking \cite{liu2024peergpt}. Additionally, Chiang et al. introduced an LLM-powered devil's advocate, which enables teams to receive opposing viewpoints during the decision-making process, which fosters critical thinking and enhances the overall quality of decisions \cite{chiang2024enhancing}.

While these collaborative activities leverage LLMs to facilitate multi-learner interaction, the scenarios explored in most studies generally do not emphasize strict time constraints and knowledge grasp and application. In contrast, as a high-paced and time-sensitive form of collaborative learning, classroom formal debate focuses more on efficient teamwork and fast group learning than the aforementioned studies. In high-intensity scenarios, the emerging Team-LLM interaction modes exhibit notable nuances compared to low-intensity, unrestricted environments. These unique nuances of the debate offer valuable insights into the potential to use AI to support such dynamic learning settings.

\subsection{Classroom Debate as Collaborative Learning}

Debate is a structured dialogue where participants present, support, and refute viewpoints on a specific topic. It fosters a deeper understanding of complex issues and develops critical thinking, communication, and teamwork skills through logical arguments and evidence. Debate, as an educational strategy, promotes active learning by guiding learners from lower-level thinking to higher-order thinking, such as analysis and evaluation \cite{vo2006debate}. It also fosters critical thinking by encouraging the exploration of diverse perspectives, improving understanding, and enhancing skills in communication and teamwork. Roy's study found that classroom debates are more effective than traditional lectures in developing higher-order thinking skills \cite{roy2005debating}. Classroom debates can be regarded as a form of collaborative learning, with related research showing that debates positively impact learners' communication and thinking skills. Brown's study indicates that debates, as a form of collaborative learning, allow students to actively participate in discussions in real classroom environments. This participation fosters student interaction and strengthens the collaborative relationship between teachers and students as they work together toward a common goal. These are all essential skills in collaborative learning \cite{brown2015use}.

In the communication process, Berdine points out that learners must temporarily set aside their biases. This practice of perspective-taking helps them become more open to different viewpoints, enhancing their ability to collaborate and communicate effectively within a team \cite{berdine1987increasing}. Additionally, Gregory et al. highlight that debates require both the audience and learners to evaluate different choices \cite{gregory2005debate}. Therefore, it aligns with the requirements for developing thinking skills by advancing through the levels of Bloom's Taxonomy \cite{kennedy2009power}. Learners can actively analyze, discuss, and apply content meaningfully rather than passively absorbing information. This active logical thinking process allows for more effective learning \cite{bonwell1991active}, thereby improving their problem-solving skills. Therefore, Brown points out that learners need to evaluate and construct arguments during the debate process. This activity not only improves learners' ability to solve complex problems but also enhances the development of their logical thinking skills \cite{brown2015use}. Furthermore, formal debates, as structured and regulated activities, are beneficial to enhance participants' rhetorical skills and their adherence to the rules of decorum \cite{vo2006debate}. In summary, the learning objectives of formal debate include gaining a deeper understanding of complex issues, enhancing critical thinking, improving communication and teamwork skills, fostering higher-order thinking such as analysis and evaluation, mastering rhetoric, and adhering to the rules of decorum.

Integrating group learning theories and AI-supported debates provides a robust framework for enhancing collaborative learning. Vygotsky's \textbf{Zone of Proximal Development} (ZPD) \cite{barrs2024zpd} underscores the importance of scaffolding, such as adult guidance, peer collaboration, or technological support, in enabling learners to surpass their independent capabilities. \textbf{Sociocultural theory}, as articulated by Mercer \cite{mercer2002words, mercer2007dialogue}, highlights the pivotal role of language in linking social interaction with cognitive development, where exploratory talk serves as a critical medium for knowledge construction. Furthermore, \textbf{Intersubjectivity}, as described by Rogoff and Wertsch \cite{rogoff1990apprenticeship, wertsch1991voices, wertsch1991sociocultural}, emphasizes the importance of shared understanding in collaborative tasks. The concept of \textbf{Sociol-cognitive Conflict}, proposed by Doise and Mugny \cite{doise1984social}, identifies cognitive adjustment through conflict and negotiation as a mechanism for deep learning. Finally, \textbf{Exploratory Talk}, as defined by Mercer \cite{mercer2002words}, is a high-quality dialogic form that enhances reasoning and collaborative skills through critical negotiation and reasoned analysis.

Debate is a language-mediated thought process that can seamlessly integrate with LLMs, which also function through language. As LLM technology advances, it shows great potential in enhancing language-based learning activities across both formal and informal educational settings, such as creative writing \cite{goldi2024intelligent,mejia2024enhancing,tu2024augmenting}, public speaking \cite{park2023audilens,follmer2023adjunct}, second language acquisition \cite{zhang2024mathemyths,kim2024exploring,higasa2024keep,ma2024teach}, and reading \cite{hanel2024towards,xu2023rosita}. However, LLM-based systems have not yet been widely applied for debate assistance, with only a few studies exploring how LLMs can support learners during the debate preparation phase \cite{xu2023rosita,guo2023effects,li2024can}. Nevertheless, considering the advantages of LLMs in processing and generating language, we believe that applying LLM technology in classroom debates can significantly benefit a wide range of educational and learning contexts and collaborative learning. For example, in-classroom peer reviews are time-sensitive (during class sessions) and require participants to engage in comprehensive and critical evaluations of others' writing. This process demands thorough consideration and necessitates providing academically rigorous feedback \cite{mangelsdorf1992peer}. Similarly, Socratic seminars emphasize the cultivation of common values and free inquiry. Participants in a debate are expected to establish a foundation of shared principles (common values), which fosters a unified stance. At the same time, the spirit of free inquiry encourages participants to engage in critical thinking and dialectical reasoning. This involves collaboratively brainstorming arguments, exploring diverse cases, and anticipating counterarguments that might arise during the debate. Likewise, other learning contexts such as collaborative augmentation \cite{chinn2013learning} and problem-based learning \cite{wood2003problem} also align seamlessly with the objectives of this research on AI-infused classroom debates, which aim to promote critical analysis, cooperative argumentation, and the development of shared understanding in time-sensitive educational settings, which are areas that traditional research on collaborative learning has paid relatively little attention to. In human-AI interaction research, classroom debates, as a structured dialogue and educational strategy, offer a unique context for human-AI collaborative learning. The fast pace, high intensity, and need for interpersonal collaboration in debates highlight their educational value. Our study aims to address a gap by exploring AI's role in enhancing collaborative learning.

\section{METHODOLOGY}\label{sec3}

\subsection{Study Setup}

\begin{figure*}[htbp]
  \centering
  \includegraphics[width=\textwidth]{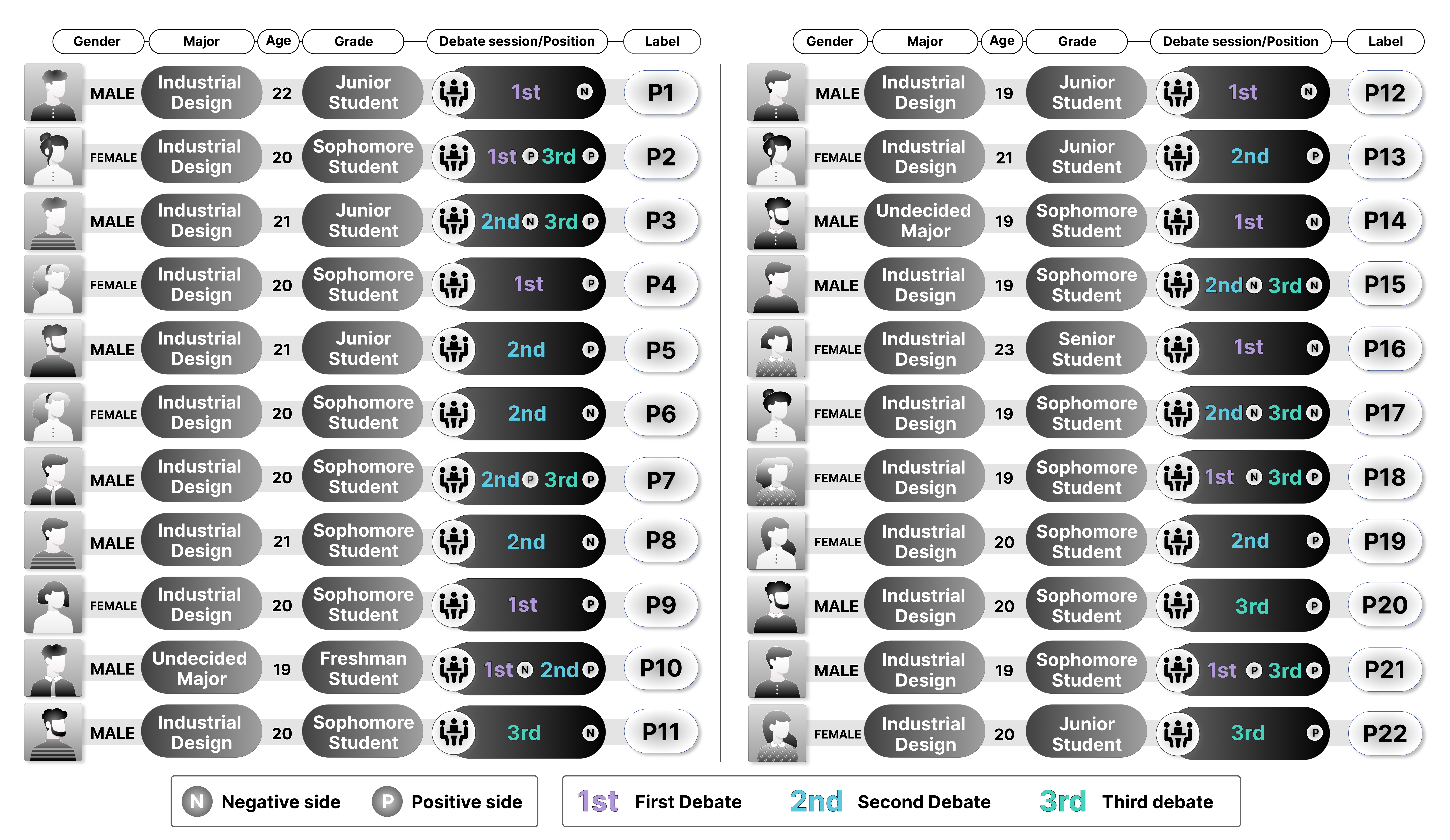} 
  \caption{\centering Information of Participants In the Classroom Debate}
  \label{fig3}
\end{figure*}

The study was conducted in the Design History course at the School of Design, Southern University of Science and Technology (SUSTech), where all the participants were enrolled. The course adopted an active learning approach, and classroom debates had been originally planned to facilitate a cooperative and learner-driven environment \cite{slavin1996research}. Initially, potential participants who were students enrolled in the Design History course were contacted to verify their willingness to engage in the study. As a result, 22 students participated, composed of 11 females and 11 males. As shown in Fig. \ref{fig3}, the distribution of their academic years was 1 freshman, 20 sophomores, and 1 senior. Participants rated their level of previous experience with using Large Language Models (LLMs) at 3.50 (\textit{SD} = 0.72) on a 1 to 5 scale (no experience to very experienced). Detailed questionnaire and survey results are shown in Appendix \ref{appendix_2}.

Three rounds of five-to-five debates were conducted over four weeks, with all participants present in each round. Each round of debates took approximately 1.5 hours. The participants were randomly assigned to either the debate's affirmative or negative side, and each side consisted of five team members. As the audience, other participants were tasked with observing the debates and voting at the beginning and end, and audience members also helped with timekeeping. The teams were able to use their digital devices during the debate. The rules stipulate that only one laptop can be accessed and use ChatGPT 3.5 in a group with no additional restrictions. By restricting AI usage to a single device, we encourage participants to develop their collaborative strategies autonomously rather than solely using AI. Rich data were gathered through classroom observations and interviews, aimed at deepening the understanding of participants' metacognitive processes (i.e., assessment and monitoring of their cognitive operations in the AI-assisted debates \cite{mahdavi2014overview}) and their experience of teamwork dynamics. The debate process was designed to fit the classroom environment, structured into the following stages (see Fig. \ref{fig2}):

\begin{figure*}[htbp]
  \centering
  \includegraphics[width=\textwidth]{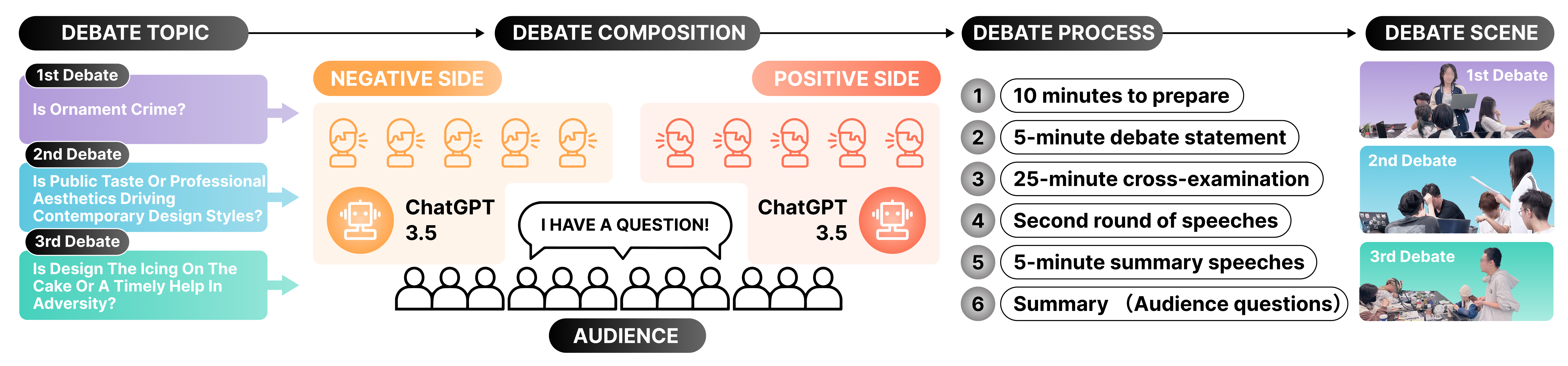} 
  \caption{\centering Overview of Human-AI Collaboration Learning Debate Scene and Process}
  \label{fig2}
\end{figure*}

\begin{enumerate}
    \item \textbf{Topic Announcement and Grouping}: The debate topic is introduced, and teams are formed.
    \item \textbf{Debate}: This includes 10 minutes of preparation time, 5 minutes for presenting initial arguments, and 25 minutes for cross-examination, followed by another round of 5-minute responses.
    \item \textbf{Team Summary}: Each team has 5 minutes for concluding remarks.
    \item \textbf{Whole-class Discussion and Wrapping-up}: After the team summary speeches, an \textit{"Audience Questions"} session took place, where the audience pose questions. All students were encouraged to engage in the discussion, enhancing the collective learning atmosphere \cite{kasneci2023chatgpt}.
\end{enumerate}

The authors' institution reviewed and approved this study (IRB No.: 20240274). All the participants gave informed consent, and the research data was anonymized and used only for academic purposes.

\subsection{The Topics of Debates}

\subsubsection{Debate 1: Is Ornament Crime?}

The first debate centered on Adolf Loos's \textit{"Ornament and Crime,"} lasting approximately 96 minutes. This debate explored whether the ornamental elements in modern design lead to rapid obsolescence and resource waste. In addition, it examined the relevance of the notion in the context of contemporary society's diverse needs and cultural expressions. This debate topic allows students to probe the significance and function of ornamentation in different cultures. The discussion focused on the contrasting roles of ornamentation in modern versus traditional societies. It sought to find a balance between the pursuit of functionalism and the preservation of cultural expression, comprehensively understanding the role of ornamentation in modern design.

\subsubsection{Debate 2: Is Public Taste or Professional Aesthetics Driving Contemporary Design Styles?}

The second debate lasted approximately 85 minutes and occurred about two weeks after the first. This debate explored whether contemporary design styles are predominantly shaped by the preferences and demands of the public or by the professional aesthetic principles upheld by designers, critics, and scholars. This topic enables students to explore how aesthetic standards are formed, how design trends emerge, and the intricate relationship between the acceptance and consumption of designs. Through this discussion, students investigated how public and professional aesthetics influence design practice and market trends, thereby understanding their interactions and impacts.

\subsubsection{Debate 3: Is Design the Icing on The Cake or a Timely Help in Adversity?}

\begin{figure*}[htbp]
  \centering
  \includegraphics[width=\textwidth]{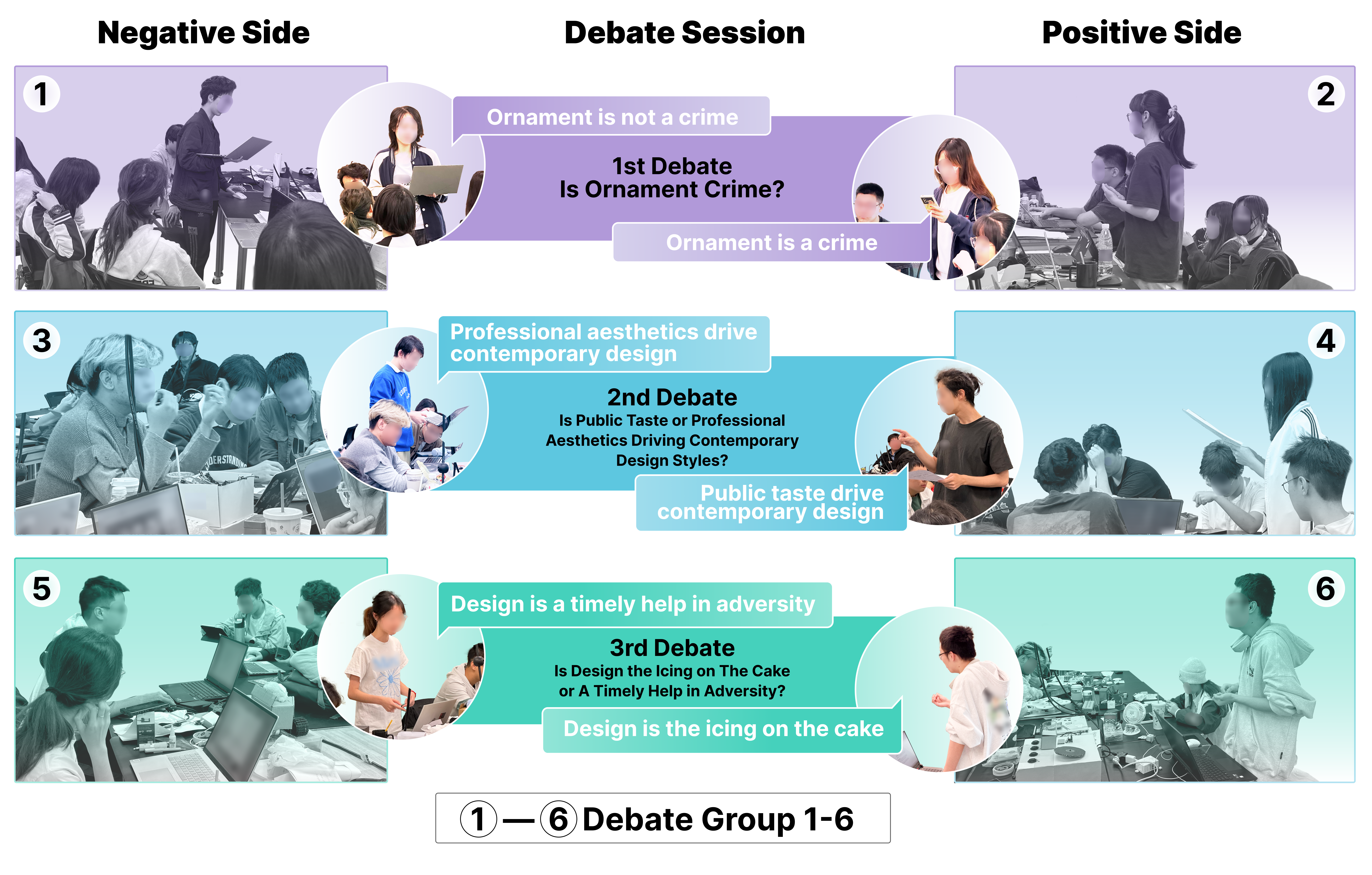} 
  \caption{\centering Three Debate Processes and Debate Topics}
  \label{fig4}
\end{figure*}

The third debate lasted approximately 90 minutes, about three weeks after the first. This debate explored the role of design in modern society: is it merely an enhancement of visual appeal and product value, or does it have the capacity to address substantial problems and even reshape society? This topic enables students to investigate and comprehend the dual functionality of design in terms of aesthetics and practicality, as well as how design exerts its unique influence in various contexts. Through this discussion, students examined the balance between enhancing consumer experiences and solving core societal issues, understanding the potential and hurdles of design in promoting social and economic development (see Fig. \ref{fig4}).

\subsection{Data Collection and Analysis}

All the debate sessions were recorded, with a teaching assistant responsible for data collection and note-taking. The participants who gave consent were included in the recordings. The teaching assistant also provided operational support, such as clarifying debate rules or responding to technical issues. After debates, semi-structured interviews were conducted with all participants individually, and they were encouraged to discuss their personal experiences. The structured questions included how students interacted with the AI tool and each other (including teammates and opponents) and how they experienced these interactions. Based on their responses, more specific and detailed questions might be asked. The average interview duration for each participant was 30.75 minutes. The total duration of audio data collected in this study was 1022.15 minutes, including recordings from all participating students. All audio data were transcribed and manually proofread, totaling 370,551 words. 

First, two coders independently coded the data's initial coding, followed by a review by a third researcher to ensure consistency and completeness. In cases of disagreement during the coding process, discussions were held to reach a consensus. For the thematic analysis \cite{braun2012thematic}, researchers annotated the transcribed text based on audio recordings to identify key terms, critical concepts, and collaborative behaviors. Audio recordings were instrumental in confirming certain behaviors and emotional responses, allowing for a deeper understanding of the behavioral patterns observed during the debates. Subsequently, the data were coded and clustered into hierarchical themes. Through iterative analysis and discussion, we distilled 10 primary themes and 29 sub-themes that reflect participants' shared experiences and unique perspectives. Each theme is supported by multiple representative quotes to ensure the results' representativeness, showcasing the broader contextual significance of the sub-themes. To gain broader insights, we incorporated classroom recordings and dialogue transcripts between groups and ChatGPT as supplementary materials. These materials provided contextual examples to support the thematic analysis results, enhancing the understanding of the findings in specific contexts and validating and refining certain themes and patterns.

In three debate rounds, participants sent 59 messages, totaling 8,181 words, and an equal number of GPT responses, totaling 37,752 words. On average, each user message contains 139 words (\textit{SD} = 322), while each GPT response averages 640 words (\textit{SD} = 314). Detailed statistics are shown in the Appendix \ref{appendix_1}.

\section{FINDINGS 1: EMERGENT WAYS LEARNERS COLLABORATE WITH AI IN CLASSROOM DEBATES (RQ-1)} \label{findings1}

In this section, we analyze users' behavior during collaboration with AI to gain insights into how teams interact with AI. This includes examining how individuals pose questions to AI and utilize AI, the division of labor within the group on how to use AI, and integrating other tools that assist collaboration.

\subsection{Ways Learners Pose Questions to AI}\label{pose}

During the debate, when learners collaborate with AI in a team setting, they ask questions differently depending on the situation, seeking more targeted answers. This study identified three distinct approaches to questioning AI.

\subsubsection{Approach 1: Learners provide keywords to AI and instruct it to form complete answers.}

For example, P(7) mentioned that when they want AI to find examples to counter an opponent's argument, they first offer more specific keywords such as \textit{"1. Environmental degradation, 2. The shift in social values, 3. Vanity, 4. Class division, 5. Increased living costs, 6. Homogenization, 7. Erosion of cultural styles, 8. The shifting communication expectations influenced by ornaments}." Alternatively, they might ask the AI to \textit{"generate examples."} P(7) also noted, \textit{"We want to find an example where public taste influenced design in a particular field,"} and thus asked the AI to provide an example of \textit{"a company undergoing design changes driven by public taste"} to support their argument. This questioning method often occurs when the user has already formed a preliminary debate strategy and uses specific prompts to guide the AI's response.

\subsubsection{Approach 2: Learners provide background information to AI to help it form appropriate responses}

P(11) noted that when discussing the topic \textit{"ornament is a crime"}, he would \textit{"first consider different scenarios, such as the periods before and after World War II, where perspectives would differ. After World War II, modernist ornament became more minimalist and utilitarian. However, if we go back to the 1920s, during the Art Deco era before the Great Depression, this viewpoint might have been seen as unconventional."} He organized the background information and input into AI. Additionally, participants provided character background information, allowing the AI to respond in a given context. 

P(21) explained, \textit{"For a particular historical figure I'm interested in, I can have the AI role-play, simulating conversations with that character."} By offering detailed historical contexts, scenario settings, and character backgrounds, participants guided the AI to simulate viewpoints and attitudes from specific periods or figures more accurately, enhancing the authenticity and persuasiveness of the AI's responses.

\subsubsection{Approach 3: Learners provide opposing viewpoints to AI and ask for rebuttal strategies}

For instance, P(16) directly asked the AI, \textit{"How should we define 'icing on the cake' versus 'a timely help' in a more advantageous way?"} In another situation, participants instructed the AI to generate rebuttals, such as when P(16) asked the AI, \textit{"Please refute the claim: accessories like necklaces and bracelets create inequality for people with disabilities,"} or when P(21) requested the AI to \textit{"write a paragraph rebutting the opponent's argument based on their stated points."} This method is common during free debates, where users promptly respond to an opponent's arguments.

\subsection{Ways Individual Learner Utilizes AI-Generated Content} \label{learner utilize}

When faced with AI-generated content, debate participants adopt various methods to filter, discuss, and refine the information to improve their own viewpoints and build consensus among group members. This section elaborates on four primary ways in which AI-generated responses are utilized. This study identified four individual filtering behaviors when analyzing participants' interactions with AI-generated content. 

\subsubsection{Mode 1: Learner uses appropriate AI responses directly as arguments}

For example, P(16) mentioned, \textit{"I select responses from the AI and use them for rebuttal,"} and P(4) noted, \textit{"I take some descriptions from AI-generated content for certain parts of the debate."} P(2) also commented that AI has a strong logical flow and its answers \textit{"make the overall argument more coherent and logical... When summarizing, I use AI responses to present our final, strongest argument."} 

\subsubsection{Mode 2: Learner engages in deeper thinking and filters AI responses to develop their own perspectives}

For instance, P(12) said, \textit{"I generate some ideas based on debate keywords, select a few valid points from the AI's responses, and then use them to form my arguments."} He emphasized that his core argument must be derived from AI-generated content through personal understanding, reflection, and transformation. P(6) also stated, \textit{"When we need an appropriate example quickly, we still have to summarize and filter AI responses to develop our understanding."} 

\subsubsection{Mode 3: Learner explores additional cases to enhance credibility}

For instance, P(9) noted, \textit{"AI provided many arguments, and I selected some, but I also found more real-life examples to reinforce them before countering the opposition."} P(17) added, \textit{"I use an AI-generated example, but I also think through or search for additional cases to support it."} 

\subsubsection{Mode 4: Learner asks AI to explain the filtered content further}

P(2) mentioned that AI provided multiple responses, and after adopting some of them, they would ask AI to \textit{"explain them in more detail to better understand them."} Similarly, P(20) stated that after reiterating their question several times to the AI, they eventually received a satisfactory answer. Through filtering, participants can extract the most effective parts of AI-generated content, incorporate their understanding, and supplement or refine the information. Repeated inquiries can also result in more complete arguments and viewpoints.

\subsection{Ways Group Members Interact with AI-Generated Content} \label{group interact}

Once AI provides responses, group members often discuss the content, which can be categorized into two types of group interactions. 

\subsubsection{Mode 1: AI-generated responses help build consensus among group members}

For instance, P(5) mentioned, \textit{"We subconsciously believe that AI's responses make sense, and AI's answers help reduce effort by resolving some disagreements within the group."} In such cases, AI-generated content helps bridge gaps between group members, fostering consensus and enhancing the efficiency of discussions and decision-making. 

\subsubsection{Mode 2: Group members classify AI-generated responses to provide a basis for more in-depth discussions}

P(12) explained, \textit{"We retain the relevant responses and discard the off-topic ones, then discuss them further. The primary purpose of using AI is still to foster group discussion." } Moreover, P(10) said, \textit{"When AI provides an opinion, we don't treat it as an expert. We first critique it, and this gradually strengthens our own views."} In this context, group members use AI content as inspiration and a starting point for discussion, sparking further thinking and collaborative reflection.

\subsection{Ways Learners Divide Labor in AI Usage within a Group} \label{labor}

In our study during the debate, we observed that group members developed various roles to utilize AI effectively. Some members became \textbf{\textit{AI users}}, who mediate between the group and the AI. This mediation involves directly reading the AI's responses or sharing them via communication apps like WeChat. Most members tended to act as \textbf{\textit{information gatherers}}, sharing the information they collected by posting it in group chats or verbally presenting their thoughts. These members focused on critical thinking without interacting directly with the AI. Some members opted to serve as \textbf{\textit{content evaluators}} after receiving AI-generated responses. Their role involved filtering and organizing the responses to enhance efficiency and minimize redundancy or confusion. Some members became \textbf{\textit{ad-hoc taskers}}, flexibly taking on tasks as required by the group's needs. Group members gradually developed a collaborative mode that best suited their team dynamics. Below is a summary of these roles, along with examples.

\subsubsection{Role 1: AI user}

One common role is the group's AI user, who mediates between the AI and the group members. As P(1) described, the AI user plays a dual role: driving the discussion forward and acting as a \textit{"mediator"} for information. The AI user is responsible for \textit{"inputting group members' questions into the AI without bias and translating subjective expressions into more objective ones for the AI to process."} However, as P(1) noted, this role can also be challenging, as \textit{"while I'm thinking about the AI's response, group members' questions often disrupt my thought process."} Similarly, P(7) mentioned, \textit{"Whenever there's a question, it's passed to me, and I ask the AI,"} while P(12) stated, \textit{"Group members tell me what they need to ask, and I input them into the AI, then share the AI's response in the group chat via WeChat group."} 

Some AI users also subjectively process the information before sharing it with the group. For instance, P(4) said, \textit{"The group's AI user will filter the AI's responses and send relevant phrases such as 'ornament in modern design plays multiple roles' to the group."} P(16) even noted that the AI user might \textit{"incorporate his own ideas, integrating multiple arguments into his statements and presenting the information in his own words."} Thus, AI users don't merely act as repeaters but filter, add, and delete content during transmission. 

Additionally, several participants noted that the AI user is often the group's focal point. As P(20) remarked, \textit{"Team members start to rely on me more,"} while P(21) observed that the AI user exhibited more active participation, especially during the debate, where they frequently stood to speak. P(17) said, \textit{"I feel like they know the most in the group because they know the AI's responses, everyone's arguments, and how everything is integrated; they also provide feedback to the AI."} This indicates that AI users tend to become key figures in the debate.

\subsubsection{Role 2: Information gatherer}

Another role is the group's information gatherer, who collects case information for the group discussion. As P(12) mentioned, \textit{"Some members take note of key points from the discussion and search for relevant materials."} P(6) added, \textit{"I'm responsible for finding relevant examples for debate points. If they think the examples are useful, they can use them."}

\subsubsection{Role 3: Content evaluator}

The group's content evaluator summarizes and assesses collected content. As P(1) pointed out, debaters responsible for opening or concluding the debate must cover all discussion points so they take notes and evaluate important information to ensure nothing is missed. 

Since relying solely on human memory may result in the omission of key points, they usually take notes: \textit{"The content evaluator is our final debater; his role is not to engage in early debate but to take notes and summarize everything at the end."} The people (content evaluators) may also use AI to integrate different ideas from group members. P(13) stated, \textit{"I usually input all ideas into the AI so that when we summarize at the end, AI can incorporate our group's ideas."} Thus, content evaluators ensure that all key points are accounted for and systematically summarized during group discussions. 

\subsubsection{Role 4: Ad-hoc tasker}

Lastly, the group's ad-hoc tasker involves members taking on specific tasks as needed, especially in time-sensitive situations. P(6) said, \textit{"In the early stages of research, I sometimes find an important point. I'll respond immediately if someone brings it up during the debate."} This flexibility allows them to respond effectively to opposing arguments, using information gathered during counter-arguments preparation. P(21) noted that during the questioning phase, when time is limited, \textit{"one person operates the computer and talks to ChatGPT, another summarizes and extracts keywords, while the remaining three take on roles for summarizing, rebuttal, and offering supplementary explanations during the free debate phase. Each person can stand and speak based on their own views, flexibly responding to new ideas in the discussion."} This shows that every group member has a clear role and can efficiently handle information and respond to debate points.

\vspace{1em}

\noindent Overall, the optimal use of AI by different groups varied depending on their specific circumstances. Some groups clearly defined roles, where responsibilities were fixed. In high-intensity, time-constrained classroom debate scenarios, such explicit role differentiation boosts team coordination and debate performance. In contrast, some groups did not establish fixed roles, opting for flexible role adjustments. They relied more on interpersonal interactions and manual summarization. While this flexibility accommodated individual preferences, it often required additional coordination under the pressure of a fast-paced debate to achieve optimal efficiency. Overall, explicit role differentiation appeared better suited for high-paced debate scenarios, contributing to improved group debate performance.

\subsection{Ways Learners Record, Exchange and Search Information beyond AI} \label{record and search}

In group collaboration, teams often use various tools in addition to the designated AI tools to record ideas that come up during their work. This section explains how groups exchange and transmit information beyond AI tools.

\subsubsection{Approach 1: Learners use social media to search and exchange information}

First, our study found that group members frequently use communication apps like WeChat to exchange information and social media like Zhihu or RedNote to search for information. For instance, P(20) noted, \textit{"I copied and pasted all AI responses into the WeChat group, while other members used different search engines, like Zhihu and RedNote, to find relevant information. We set up a WeChat group and shared all the materials we found."} P(17) added that she would \textit{"take screenshots of collected information and then post it in the WeChat group."} Some participants also use shared mind-mapping apps for brainstorming. For example, P(12) stated, \textit{"We used tools like mind maps in Mubu to organize arguments. I would copy AI responses and material found by other members from Zhihu or RedNote and organize them under the corresponding argument. This way, when I need to speak, I can directly extract and present the information."}

\subsubsection{Approach 2: Learners use traditional and digital notes to record ideas}

Additionally, timely written records are commonly used to record ideas, including traditional pen-and-paper notes and digital notes. P(5) mentioned using a notebook to record key points and then providing them to the AI for language refinement. After recording, P(5) often uses paper and pen to note ideas during the opponent's speech. P(20) explained that he would \textit{"provide AI with several brief words or phrases from these notes to refine the rebuttal language."} He then asks AI to \textit{"further generate responses based on the handwritten key points."} P(12) said she also uses the \textit{"memo app on her phone,"} while P(15) mentioned using a \textit{"note-taking app on a tablet to record information."}

\subsubsection{Approach 3: Learners exchange information via face-to-face communication}

Moreover, when time is limited, participants tend to choose face-to-face communication. P(4) mentioned, \textit{"AI first proposed a point, and we naturally started discussing it. We collectively accepted AI's viewpoint during the discussion and clarified the argument."} In such time-sensitive situations, face-to-face communication allows for more fluid and rapid exchanges, facilitating a deeper exploration of the ideas presented.

\section{FINDINGS 2: THE ADVANTAGES AND RISKS OF AI IN FAST-PACED COLLABORATIVE LEARNING (RQ-2)}

In this section, we analyze participants' behavior and feedback to explore the role of AI in fast-paced, collaborative learning environments. Specifically, we examine how AI reduces social anxiety, facilitates effective communication among participants, lowers the barriers for novice debaters, stimulates individual critical thinking, and offers diverse perspectives that provoke discussion among team members.

\subsection{Advantage 1: Reducing Social Anxiety and Communication Barriers Among Learners} \label{reduce social}

We observed that AI's involvement significantly alleviates participants' social anxiety in collaborative environments. In team collaboration, AI helps break discussion deadlocks and facilitates communication among group members. Specifically, AI serves as a conversational partner through a convenient, low-pressure interaction mode, reducing users' social anxiety.

\subsubsection{Reducing the Social Anxiety of Asking Questions}

The first observation is that AI helps to \textit{"reduce the social anxiety of asking questions due to the fear of bothering others."} For example, P(12) mentioned that he was hesitant to ask teachers for help, stating, \textit{"I don't dare to ask teacher because I don't want to bother him,"} but with AI, \textit{"I don't feel much social anxiety."} Similarly, P(11) noted that when faced with many questions, \textit{"I don't need to ask others; I can just ask AI."} Additionally, AI's polite and service-oriented response style also helps reduce social anxiety. P(18) commented, \textit{"AI makes you feel like you are in control, and it responds very politely, like talking to a friend."} This conversational style makes the interaction feel comfortable. 

\subsubsection{Reducing the Effort of Adjusting Language Based on the Person}

Another way AI eases communication is by eliminating the need to tailor language based on the characteristics of different individuals. P(16) mentioned that when faced with real people, his communication is influenced by the other person's personality, and he must adjust his language accordingly. In contrast, with AI, \textit{"there are fewer concerns."} P(1) added, \textit{"I feel that communicating with AI is more direct. I can ask whatever I want without considering the other person's personality. If I were talking to a real person, I might adjust my questions based on their character."} Lastly, P(20) mentioned that when AI gives an incorrect answer, he can directly ask AI to modify its response.

\subsubsection{Allowing Asking Unlimited Questions Without Restrictions}

Participants can ask AI questions as often as needed without feeling limited. P(2) noted, \textit{"When communicating with people, you may have concerns, like worrying if your question seems silly. However, with AI, you can ask anything freely, from wild ideas to any other question, without worrying about such concerns."} P(16) explained that when communicating with people, if your question isn't clear the first time, they may keep asking follow-up questions until you clarify it completely. But with AI, \textit{"you can ask questions anytime and receive immediate responses without worrying about unclear expressions."}

\subsubsection{Breaking Communication Deadlocks and Facilitating Group Discussions}

AI helps break deadlocks in discussions and promotes smooth communication among group members. First, AI can quickly generate large amounts of text to kickstart discussions within the group. For example, P(17) mentioned, \textit{"At the start of the discussion, everyone might not have specific ideas, and the group could be in a deadlock. AI generates a lot of content, and although not all of it is usable in the debate, it serves as a starting point. After seeing AI's response, our minds start working, and the discussion flows."} P(10) similarly mentioned that AI's text generation has at least two benefits: \textit{"First, it ensures you aren't left without anything to say, and second, it prevents you from having nothing to contribute at all."} 

Next, AI stimulates communication by generating topics that increase interaction. P(10) said, \textit{"AI helps fill awkward silences. When the discussion stalls, it provides absurd reasons."} While these AI responses may not be practically useful in the debate, the\textit{"imperfect"} or \textit{"absurd"} reasons can serve as a catalyst to start conversations. P(10) added, \textit{"When no one has any ideas, AI can generate some topics, and people will discuss AI's answers and develop their own ideas."} 

AI can also help group members focus on idea generation through its quick language organization. P(14) noted, \textit{"Another advantage is that AI's word choice is always on point. Sometimes we struggle to find the right words or use placeholders, but AI's words are accurate and reliable."} P(6) mentioned that AI helped them rewrite the text, which \textit{"unblocked our earlier mental blockages, allowing us to naturally continue the discussion."}

\vspace{1em}
\noindent These behaviors show that when interacting with AI, learners can reduce their social anxiety and enjoy greater freedom of expression. Communication becomes more direct and efficient by not considering the personality, background, number of questions, or other interpersonal factors. This improves communication efficiency and allows people to express their thoughts more freely without fear of being misunderstood or offending others. In this way, AI promotes smoother information exchange.

\subsection{Advantage 2: Lowering the Threshold and Providing Scaffolding for Novices}\label{scaffold}

For novices, AI lowers the entry barrier to debate by quickly constructing a debate framework in a short time. It can provide a large amount of content for beginners to reference, helping them think from multiple angles about the same issue. Novices can also quickly and accurately search for and understand terminology and background information, helping them build knowledge in relevant fields. Additionally, AI can generate professional suggestions, particularly valuable for less experienced users, helping them express and participate more effectively in professional discussions. AI also helps organize and clarify learners' original ideas, allowing for better structure and expression of their thoughts. Below, we explain these findings in more detail.

\subsubsection{Providing Novices with Much Information Quickly}

First, AI helps novices construct thought and debate frameworks by quickly providing information. For example, P(5) mentioned, \textit{"At first, we were all quite unclear about the debate process. We only knew we had to present arguments and then engage in free debate, but we didn't have a clear concept of how to structure an argument."} By using AI, \textit{"the entire process is listed for you. It tells you what the opposition might say and what you need to say, forming an argument framework."} P(4) stated, \textit{"During debate preparation, I found AI extremely helpful in organizing language quickly. It provided a clear framework, such as how to speak and what strategy to use, and also listed steps like introduction, explanation, and summary. The structure was very clear, and I could quickly understand and use these tips, making the preparation smoother."} P(6) added, \textit{"Since we didn't have much time, we couldn't spend days thinking about arguments and crafting polished statements. AI gave us an outline, and we selected the most suitable content, greatly reducing our preparation time."}

\subsubsection{Helping Novices Better Organize and Express Their Ideas}

In debates, novice participants mentioned that AI helped them organize and structure their thoughts. For instance, P(3) said, \textit{"I gave AI some of my ideas or a long description without needing to summarize it myself. Sometimes, it was just a simple description of an objective fact or a collection of random thoughts, and AI extracted the key points and organized them."} P(7) added, \textit{"We encountered problems that we couldn't structure with our current knowledge. In this case, we used AI. AI gave us a logical framework, which became the foundation for clearer discussions."} Participants also noted that AI helps them quickly generate language that conforms to debate norms. P(2) said, \textit{"I'm not very good at speaking in public, but with AI's help, I now know how to start, like 'Dear judges, esteemed opponents... First, our side believes...' I didn't know these before, but AI quickly helped me structure my points in a specific format."}

\subsection{Risk 1: Causing Information Overload in Fast-Paced and High-Intensity Collaborative Learning Environments}\label{overload}

During debates, AI provides a large amount of information that requires participants to spend time processing, leading to information overload. Due to the high intensity of debates, participants' cognitive resources become particularly valuable, making them more sensitive to information overload. Excessive or irrelevant information can disrupt participants' train of thought in such high-pressure environments.

\subsubsection{Providing Too Many Ideas}

First, participants mentioned that AI provided too many ideas, which increased the time needed for understanding. For example, P(8) stated, \textit{"I feel like AI can generate a lot of ideas in a short period, but first, we need time to process and understand, which leaves me with no time for personal thinking because AI is too efficient."} Participants also noted that during short scenarios, such as cross-examinations, \textit{"I first had to understand the content provided by AI, then supplement what was missing, which took quite a bit of time."} This demonstrates how an overabundance of AI-generated answers can lead to cognitive load \cite{sweller2011cognitive}. Users may struggle to fully develop and explore their own ideas when they are busy understanding and improving AI-generated content.

\subsubsection{Providing Irrelevant Content}

On the other hand, AI tends to provide irrelevant content, making it difficult to quickly process and filter large amounts of text. P(6) mentioned, \textit{"AI's arguments are usually a whole paragraph. Sometimes, when we want to find the answer, we must read the whole paragraph and summarize it ourselves."} This process consumes time. Additionally, AI's irrelevant points can disrupt users' subsequent thinking. P(1) mentioned, \textit{"Some ideas are pointless, so after reading them, you have to abandon them."} It is clear that some AI's irrelevant responses can lead participants to consume valuable thinking time in the high-paced debate scenario.

\subsection{Risk 2: Causing Cognitive Dependency and Limiting Individual Autonomy}\label{dependency}

This study found that many participants expressed a sense of dependency on AI, which grew over time as they increased their AI usage frequency. AI's involvement in group collaboration sometimes limits participants' autonomy. 

\subsubsection{Causing Cognitive Dependency}

Participants noted that AI's rapid and complete answer could foster cognitive dependency. For instance, P(18) said, \textit{"AI gives so many answers at once that our team skips the thinking phase."} He also pointed out, \textit{"AI gave us a long answer, and we were too lazy to summarize it ourselves."} This led the team to rely on AI's output. Although this reduces the workload, it may cause team members to miss opportunities for deeper understanding. P(8) also stated, \textit{"When AI presents the idea directly, we feel like it makes sense. AI-generated content is like a ready-made dish served to us."} This reinforces participants' tendency to focus excessively on AI's responses, reducing opportunities for personal reflection. P(6) pointed out, \textit{"Debating is more about listening to the other side and finding rebuttal opportunities. Now, we're more focused on AI's content, which diminishes the importance of listening."} This particularly affects concluding or opening statements, as P(8) stated, \textit{"What should be a collective summary from the discussion has now become reading from AI's script."} 

Additionally, using AI frameworks may create resistance to new ideas. P(1) mentioned that, after building on AI's framework, team members often feel \textit{"theoretical systems are already complete,"} which leads to a subconscious resistance to new ideas. Even potentially valuable ideas might be ignored or rejected due to these preconceived notions.

\subsubsection{Reducing Individual Autonomy}

Some participants felt that the AI-determined arguments in the debate process lowered their feelings of involvement. P(1) stated, \textit{"The arguments didn't come from me."} Participants may not feel strong ownership of these AI-generated arguments. P(8) mentioned that AI reduced the \textit{"sense of achievement that comes from going from zero to one."} P(6) echoed, \textit{"The concluding and opening speeches were almost entirely based on AI's content. While convenient and quick, it lacked personal input, and felt a bit disappointed,"} 

Additionally, participants felt that interacting with AI lacked real human interaction's stimulating and engaging experience. P(5) emphasized that AI lacks the interactive dialogue experience and emotional connection that make conversations enjoyable and meaningful. He felt that AI could not be considered a true debate participant but an auxiliary tool. He said that using AI \textit{"doesn't reflect the true meaning of debate"} since it should represent personal thoughts and opinions, which made him feel guilty.

\subsection{Risk 3: Low-Quality AI Responses Weakening Reliability of AI}

In debate activities, learners still perceive the quality of AI-generated content as low in certain instances. Below, we outline three situations where this occurs.

\subsubsection{Incomplete and Inconsistent in AI-Generated Information}

Participants noted that AI's limited text remembering and content alignment capability caused conversation content to be incomplete and inconsistent. P(8) mentioned, \textit{"AI doesn't have a long-term memory. For example, if I ask it to respond in Chinese, after a few minutes, it switches back to English."} P(18) similarly pointed out, \textit{"AI doesn't absorb everything mentioned in a conversation. The content we discussed earlier often gets forgotten after a few rounds."} He also pointed out, \textit{"Sometimes you ask AI for something very specific, like 'Who said this quote?' or 'Which era is this from?' AI sometimes gives wrong answers. So during debates, we didn't ask it much because the information it gave wasn't useful,"} P(11) explained. When AI made such mistakes, participants expressed a decline in trust. P(21) said, \textit{"Later on, we didn't trust AI as much. We relied more on our own thinking and stopped using AI as much."}

\subsubsection{Cultural Bias in AI-Generated Responses}

Participants expressed that AI's culturally specific examples could lead to confusion and perceptions of cultural bias. P(1) felt that AI provided \textit{"a very bad example"} when it referenced cultural contexts that \textit{"people weren't familiar with,"} making these concepts \textit{"relatively abstract."} Additionally, AI would sometimes mention examples such as \textit{"Apple products"} or \textit{"Air Jordan shoes"}, which P(1) believed reflected cultural bias. He noted, \textit{"Once the examples are grounded in (our) everyday life, our discussions become easier, as people can more easily understand those examples and empathize when facing real-world situations."}

\subsubsection{Formulaic Responses from AI}

Our analysis of participant conversations with AI showed that some participants felt AI's responses were formulaic. P(20) noted, \textit{"AI might respond to a particular argument with three sentences: the first is 'what,' the second is 'why,' and the third is a 'summary,' like a classic three-part structure. Every response is like that."} P(1) also stated, \textit{"AI sometimes likes to give obvious and repetitive responses."} Participants noted that AI's responses to certain questions could feel rigid and mechanical, potentially leading to a lack of diversity.

\section{DISCUSSION}

In this study, we conducted an in-depth analysis of new interaction patterns between learners and AI during fast-paced classroom debates. These interaction strategies enriched collaborative learning and revealed the benefits and risks of integrating AI into educational settings. 

In this section, we analyze and discuss the collaboration between teams and AI based on the findings and related works. We also propose several design strategies to maximize the benefits while minimizing potential risks. Additionally, we review the limitations of human-AI interfaces in supporting team collaboration and the constraints of existing AI tools in text-based interactions to explore future research directions and potential areas for improvement.

\subsection{Does AI Help or Hinder Learning Objectives in Formal Debate?}


Teamwork emphasizes the efficient achievement of goals through clear task division and effective resource coordination \cite{cohen1991teamwork}. In contrast, group learning focuses on enhancing individual cognitive abilities and constructing knowledge through collaboration. It prioritizes interaction and the development of shared understanding during the learning process, aiming for deeper and more meaningful learning outcomes \cite{driskell2018foundations}. Formal debate serves as an exemplar of both teamwork and group learning. It is a high-paced and time-sensitive scenario where the group must quickly reach a consensus or make decisions, such as formulating rebuttals against opponents' arguments within a limited time. In such contexts, group members must efficiently acquire knowledge and generate effective arguments to succeed in the debate. Ultimately, the purpose of formal debate is not only to achieve success in argumentation but also to facilitate the cognitive growth of group members through the collaborative process. In this section, we will explore whether integrating AI in formal debates helps or hinders the achievement of learning objectives.

In Section \ref{labor}, we showed how, throughout the debate process, group members naturally divide labor, each taking responsibility for specific tasks such as gathering information, utilizing AI, and evaluating content. Compared to traditional debates, introducing AI diversifies the potential roles within a group, such as assigning an "AI user" role. According to \textbf{Intersubjectivity} \cite{rogoff1990apprenticeship, wertsch1991voices, wertsch1991sociocultural}, group members with different responsibilities interact to build shared understanding, collaboratively constructing knowledge. This coordination and interaction significantly enhance teamwork skills, particularly under the time constraints of formal debate.

In Section \ref{pose} and Section \ref{record and search}, we highlighted that, unlike traditional debates, which rely on face-to-face interactions or social media discussions among group members, the integration of AI adds a new dimension, exchanging ideas with AI itself. This process strengthens interpersonal communication skills and develops the ability to communicate effectively with AI, a skill that is increasingly relevant in the modern world.

In Section \ref{reduce social}, we noted that AI can reduce social anxiety among group members, helping to break communication deadlocks and facilitating group discussions. This dynamic fosters a more inclusive environment and further accelerates the growth of learners' communication skills.



In Section \ref{scaffold}, we explored how AI provides scaffolding across various aspects of formal debate, such as outlining the debate process, crafting rebuttal strategies, and adhering to formal language and etiquette. According to \textbf{Zone of Proximal Development Theory} \cite{vygotsky1978mind}, learners achieve tasks beyond their independent capabilities through collaboration and guidance in a socially constructed learning space. This guidance traditionally comes from teachers or peers, who act as a \textit{more knowledgeable other} (MKO), or from self-directed learning through books and online searches. In this study, AI plays an unconventional yet intriguing role as an MKO. While it would not be accurate to claim that AI is more dynamic than a human collaborator, the flexibility and vast knowledge base of AI make it a unique tool for scaffolding in diverse contexts. This study shows AI’s potential to create a dynamic and inclusive learning environment, extending opportunities to learners regardless of their physical or social access to traditional resources.

In Section \ref{learner utilize}, Section \ref{group interact}, and Section \ref{scaffold}, we emphasized how AI can provide large volumes of information quickly, enabling learners to gain a more comprehensive understanding of debate topics. Learners can deepen their reasoning skills and construct knowledge collaboratively by engaging in \textbf{Exploratory Talk} \cite{mercer2002words}, which involves critical negotiation. AI contributes to this process by providing real-time feedback, such as clarifying points or supplementing evidence, thereby improving discussion quality. This integration of AI supports a deeper understanding of complex issues and strengthens learners' critical thinking skills.

However, in Section \ref{overload} and Section \ref{dependency}, we acknowledged the potential drawbacks of AI. When AI generates excessive irrelevant or inaccurate information, learners may experience information overload, which can reduce their critical engagement with AI-generated content. This, in turn, risks undermining the achievement of critical thinking as a learning objective.

\subsection{Unique Contribution of AI-infused Classroom Debate}

Our study emphasizes the uniqueness of classroom formal debate scenarios, differentiating them from prior research on AI-infused group collaboration settings. Classroom debates are characterized by strict time constraints, requiring participants to rapidly formulate logically sound arguments and respond to opponents' rebuttals. This high-paced environment necessitates a clear division of labor and collaboration efficiency, driving the emergence of team-LLM collaboration patterns (shown in Section \ref{findings1}) to ensure effective communication and decision-making under tight time pressure. Furthermore, debate participants must construct logical rebuttals and strategically employ language on the spot, requiring quick thinking and precise expression.

In contrast, AI-infused group decision-making relies more on AI to facilitate divergent thinking and generate diverse creative ideas \cite{yu2023investigating}. Debates emphasize logically cohesive arguments, linguistic precision, rhetorical finesse, and strategic application of reasoning. Additionally, the intensity of debates makes cognitive resources more precious, rendering participants highly sensitive to information overload. Excessive or irrelevant information can disrupt participants' focus and hinder their responsiveness in high-pressure environments. However, other AI-infused collaborations typically do not operate under such urgency, and they usually have more time to eliminate useless information \cite{chiang2024enhancing, yu2023investigating, liu2024peergpt}. These cognitive demands highlight the need to effectively manage and filter AI-generated information in fast-paced scenarios, presenting a critical challenge for real-time AI assistance.

Moreover, Park et al. proposed some challenges of using LLMs in higher education, and our study has solved some of them, including \textit{"decreased learning and academic integrity"} and \textit{"reduced social interaction"} \cite{park2024promise}. For the former, in the AI-infused debate, AI does not directly act as a debater. In contrast, it acts as a tutor to provide scaffolds, debate rules, and strategy and as an assistant to provide vast information for debaters. Thus, debaters avoid reducing learning and academic integrity and can use tutoring to quickly grasp debate knowledge and frameworks. This approach enables them to efficiently formulate their arguments in high-paced scenarios. For the later challenge, we intentionally stipulate that only one laptop can be accessed and use ChatGPT 3.5 in a group in our study. Our findings show that because only a few people can use ChatGPT, they developed a clear division of labor and communicate deeply face-to-face or via social media, which promotes their social interaction skills. These contributions demonstrate the success of our designed AI-infused debate.

\subsection{Limitations of Current Human-AI Tools in Dialogue-Based Group Collaboration}

Based on our findings, interactive interface design directly impacts teamwork's effectiveness in collaborative learning. The current human-AI interface exhibits several limitations that may hinder the efficiency of team collaboration. First, since AI interfaces are primarily designed for one-on-one interaction, they lack the necessary support for effective information sharing among team members in group settings. This results in obstructed information flow, increasing communication costs within the team. For instance, when team members need to reach a consensus, the absence of a centralized platform for information exchange makes communication more difficult and time-consuming. 

\begin{framed}
\noindent
\textbf{\large Design Implication 1: Improving Information Sharing and Communication Flow} \vspace{1mm}

To address the challenges of \textbf{high-paced} and \textbf{high-intensity} collaborative learning, future LLM-based systems should move beyond isolated one-on-one interactions and manual workflows like copying outputs into shared online documents. Instead, they should offer integrated, adaptive interfaces for seamless team collaboration. These interfaces could support multimodal inputs, enabling users to communicate naturally through text, voice, images, and other forms within the same platform. 

Real-time interaction and shared workspaces should replace fragmented processes, allowing team members to collaboratively co-create, discuss, and edit AI-generated outputs. By incorporating features like synchronized updates, role-based views, and tools for annotating or summarizing multimodal content, such systems can streamline communication, enhance interpersonal connections, and meet the demands of fast-paced team environments. These designs align with \textbf{Intersubjectivity} \cite{rogoff1990apprenticeship} and \textbf{Sociocultural Theory} \cite{mercer2002words}.

\end{framed}

Our study also found that AI has a limited memory for text, resulting in discontinuous conversations and frequent omissions, especially when handling complex or multi-step instructions requiring repeated inputs. Furthermore, AI often lacks a clear stance when dealing with ambiguous viewpoints. When providing examples, AI responses always mix useful information with irrelevant details, which increases the user's workload of filtering content \cite{soman2023observations}. Although the long-term memory and context-alignment of LLMs are inherently determined by the model training process or the framework that LLMs use, and numerous AI scientists are actively working to address this challenge \cite{wang2024augmenting, liu2023trustworthy}, we can adopt specific strategies to make their responses more focused and efficient without compromising the LLMs' performance.

\begin{framed}
\noindent
\textbf{\large Design Implication 2: Enhancing AI's Output Using Team-members Selected Information} \vspace{1mm}

Allow team members to mark and save AI responses they find important or satisfactory during interactions, introducing a feedback mechanism with options such as "Support," "Oppose," and "Neutral." Add interface options for users to specify the stance the AI should adopt, enabling the AI to generate responses that better align with user expectations. This approach allows the AI to reference saved information in subsequent conversations, maintaining continuity in dialogue. Establish a team-level memory repository to consolidate meaningful responses saved by all members, ensuring that all key information is considered during team interactions.

Additionally, it should provide convenient search and reference functions in the interface, allowing team members and the AI to quickly locate and cite previously saved content, reducing information omissions and minimizing the need for repetitive input.

\end{framed}

\subsection{Convenience of Information Access v.s. Information Overload and Quality Issues: A Double-Edged Sword}

According to our findings, many participants indicated that AI tools allowed them to quickly obtain large amounts of information, greatly facilitating the retrieval of relevant data when needed. However, the volume of information AI provides also comes with the potential risk of information overload. Participants reported spending significant time trying to discern valuable information, and some resorted to \textit{"giving up"} by completely trusting AI's responses. This contradicts one of the key characteristics of human-AI collaborative learning: \textit{"mutual understanding"} \cite{huang2019human}. At the same time, participants acknowledged that AI's ability to provide large amounts of content quickly offered beginners an overview and helped them build a preliminary understanding. This highlights AI's dual nature in terms of information quality.

\begin{framed}
\noindent
\textbf{\large Design Implication 3: Mitigating Information Overload} \vspace{1mm}

Providing contextually appropriate explanations based on the stage of interaction allows users to extract more relevant insights and fosters deeper collaboration. For example, before responding, a system could offer preset options such as "Concise," "Moderate," or "Detailed." Users could then select the desired level of detail according to their needs, avoiding situations where they must wait for the AI to return an excessive amount of information and subsequently summarize it themselves or request the AI to summarize it again.

When addressing complex problems, AI should be designed to break down the issue into several sub-questions, progressing from simple to complex. This step-by-step approach would guide users in thinking through more intricate aspects of the problem while avoiding the overwhelming presentation of excessive information all at once. Such a design ensures that users can better grasp the key points of the response without being distracted by an overload of details, following the principles of \textit{"mutual understanding"} and \textit{"mutual growth"} in human-AI interactions \cite{huang2019human}.

\end{framed}

\subsection{Moderate AI Dependency in Group Collaboration}

Through analyzing participant feedback, this study found that the convenience and rapid information-providing capabilities of AI tools could sometimes lead to over-reliance, reducing participants' independent thinking. In time-constrained situations, some participants chose to think independently before turning to AI, reflecting their awareness of the dangers of over-reliance on AI tools and a desire to maintain their capacity for independent thought. We also noted that participants' \textit{"self-awareness"} behavior is related to their confidence levels, which echoes Chong et al.'s research, demonstrating that an individual's confidence level influences their decision to accept or reject AI suggestions \cite{chong2022human}. Specifically, individuals with higher confidence in their judgment are more likely to question or ignore AI recommendations, while those with lower confidence may over-rely on incorrect AI suggestions. This phenomenon is referred to as \textit{"human self-confidence calibration"} \cite{ma2024you}, which suggests that researchers should assess and calibrate users' confidence levels before they start using AI tools. 

\begin{framed}
\noindent
\textbf{\large Design Implication 4: Dynamic Adjustment of AI's Role in Team Collaboration} \vspace{1mm}

Future AI design should incorporate mechanisms to encourage critical thinking, particularly by adjusting AI engagement based on users' confidence levels. For highly confident users, the system can pose challenging questions or case analyses to stimulate critical thought, while for less confident users, a gradual approach with guidance and encouragement can help improve their decision-making and problem-solving abilities. This approach fosters a balance between AI assistance and independent thinking.

In addition, dynamic AI adjustment features can be integrated into team collaboration, allowing tasks and roles to be flexibly assigned based on each team member's performance and needs. Adjustments can be based on members' confidence, interaction frequency, contributions, and feedback, ensuring everyone actively engages in tasks aligned with their abilities and interests, ultimately deriving fulfillment from the collaborative process. This method balances individual participation in group efforts and enhances a sense of achievement in collaborative learning.

\end{framed}

\subsection{Limitations and Future Work}

This research has revealed several limitations during the study process. Firstly, our results may not fully represent a wide range of collaborative learning scenarios due to a limited sample size. Additionally, technological maturity has not yet fully met educational needs. The dialogue systems utilized in this study have not been widely implemented in practical study environments, and both the technical maturity and user adaptability remain to be verified. This limitation might impact the practical applicability of our findings. Data security in human-AI interaction is another significant concern. Existing studies have identified biases and stereotypes in the outputs of Large Language Models (LLMs) \cite{bai2024measuring, gupta2023bias}. Furthermore, our study lacks diversity since our participants were primarily from specific age and academic groups. The classroom types explored may also limit the applicability of our findings across broader educational formats. The absence of long-term tracking of technology interventions makes it challenging to assess their enduring impact, and the potential novelty effect introduced by technologies like ChatGPT must also be addressed in future work \cite{kim2024exploring}. ChatGPT-4 had just been released during the classroom debate experiments and was relatively slow, making it unsuitable for high-paced debate settings. Faster-response large language models (LLMs) like GPT-4o had not yet been released. Thus, ChatGPT-3.5 was chosen for its faster response times. While the focus of the study was not on the performance of LLMs but rather on the interaction patterns within AI-infused team collaboration, it is undeniable that the capabilities of LLMs could influence the information-gathering process and the degree of response hallucination \cite{huang2023survey}, ultimately impacting the outcomes of the debates. Future research could build upon this work by employing state-of-the-art LLMs to explore these dynamics further. Additionally, all participants are Chinese, which means they operate within a specific cultural context when gathering information on platforms like Zhihu and RedNote. This cultural specificity similarly influences their use of AI. Consequently, addressing how different cultural norms and practices in information retrieval can be incorporated into AI-infused collaborative learning represents a limitation of this study and a significant avenue for future research. In the future, a broader range of cultural norms needs to be included in the research, thereby improving the generalizability of the findings. Finally, although classroom debate activities were already part of the course's regular teaching practices, with occasional invitations extended to external teachers or students as observers, the setup in this study does not differ significantly from the usual classroom environment. However, we cannot rule out the potential influence of the Hawthorne effect \cite{adair1984hawthorne}, a common challenge in observational research.

\section{CONCLUSION}
This study investigates interactive behaviors in human-AI hybrid collaborative learning by creating five-to-five classroom debates involving 22 students. The findings indicate various collaborative modes emerged between team members and AI during the interaction. AI's involvement altered how humans posed questions and utilized AI-generated answers, fostering a division of labor within teams. Additionally, different methods of using tools for recording and exchanging views were observed. These processes helped groups build consensus, filter information, deepen understanding, and form diverse perspectives, enhancing team collaboration. Moreover, AI significantly reduced users' psychological burden and communication barriers within groups, lowered the entry threshold for novice debaters—such as addressing cognitive biases—stimulating individual thinking, and sparked discussions between participants. However, the study also reveals several challenges in human-AI collaborative learning. AI may lead to information overload, induce dependency, limit human autonomy, and cause the perceived low quality of AI-generated content. This leads to cognitive inertia, weakens personal initiative, and restricts the generation of new ideas. To maximize the benefits of AI while mitigating associated risks, the study proposes design strategies, analyzes the limitations of current human-computer interfaces in collaborative learning, and discusses the text memory constraints of existing AI tools in dialogue tasks. Furthermore, it highlights the impact of AI-driven task allocation on participants' motivation and engagement in collaborative learning. The study's limitations include a small sample size, insufficient technological maturity, data security concerns, the homogeneity of the user group, and a lack of long-term impact assessment. These factors provide multiple dimensions for the expansion and deepening of future research.

\section*{ACKNOWLEDGMENTS}
We sincerely appreciate the involvement of all the volunteers who participated in this study and the 22 students who contributed to the project experiments for their efforts and support in data collection and conducting interviews throughout the research. We also greatly thank the reviewers for their invaluable feedback and suggestions. This work is supported by the National Natural Science Foundation of China (NSFC) Youth Grant (Grant No. 62307024).

\bibliographystyle{ACM-Reference-Format}
\bibliography{reference}

\appendix

\section{Survey on the Use of Large Language Models (LLMs)}\label{appendix_2}

\subsection{Full Questionnaire}

\begin{enumerate}    
    \item \textbf{(Before the debate) How frequently do you use LLMs?}
    \begin{itemize}
        \item Never
        \item Rarely
        \item A few times a month
        \item A few times a week
        \item Daily
    \end{itemize}
    
    \item \textbf{In which scenarios do you primarily use LLMs? (Multiple choice)}
    \begin{itemize}
        \item Academic research (e.g., writing papers, analyzing literature)
        \item Learning assistance (e.g., solving problems, writing suggestions)
        \item Programming and technical support
        \item Entertainment (e.g., chatting, generating stories)
        \item Other: \_\_\_\_\_\_\_\_\_\_\_\_\_\_\_\_\_
    \end{itemize}
   
    \item \textbf{Do you have experience in using LLMs and capable of effectively using them to accomplish tasks?}
    \begin{itemize}
        \item Strongly disagree - 1
        \item Disagree - 2
        \item Neutral - 3
        \item Agree - 4
        \item Strongly agree - 5
    \end{itemize}
    
\end{enumerate}

\subsection{Results}

\begin{figure*}[htbp]
  \centering
  \includegraphics[width=0.5\textwidth]{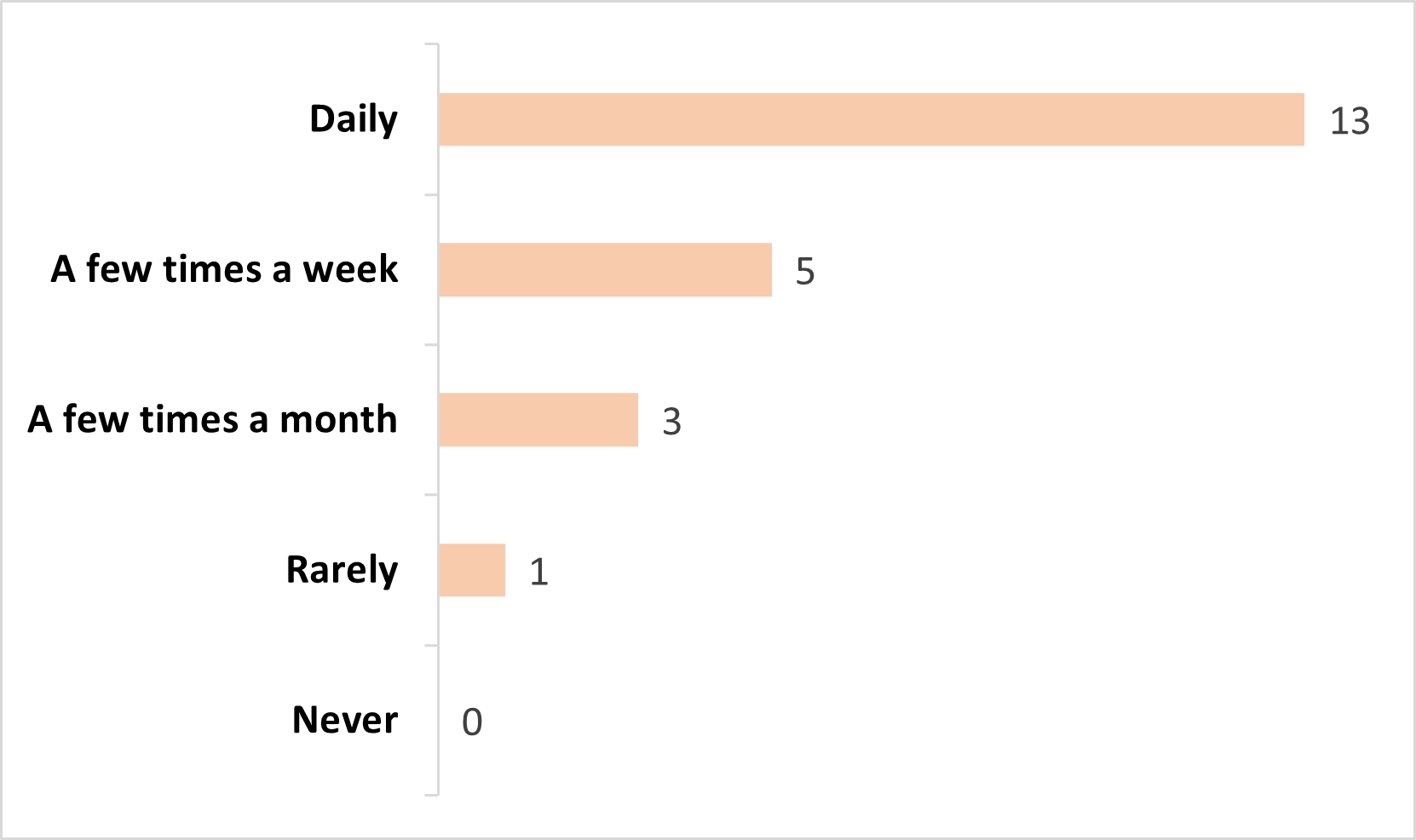} 
  \caption{\centering Results for Question (1): \textit{"(Before the debate) How frequently do you use LLMs?"}}
  \label{fig5}
\end{figure*}

\begin{figure*}[htbp]
  \centering
  \includegraphics[width=0.5\textwidth]{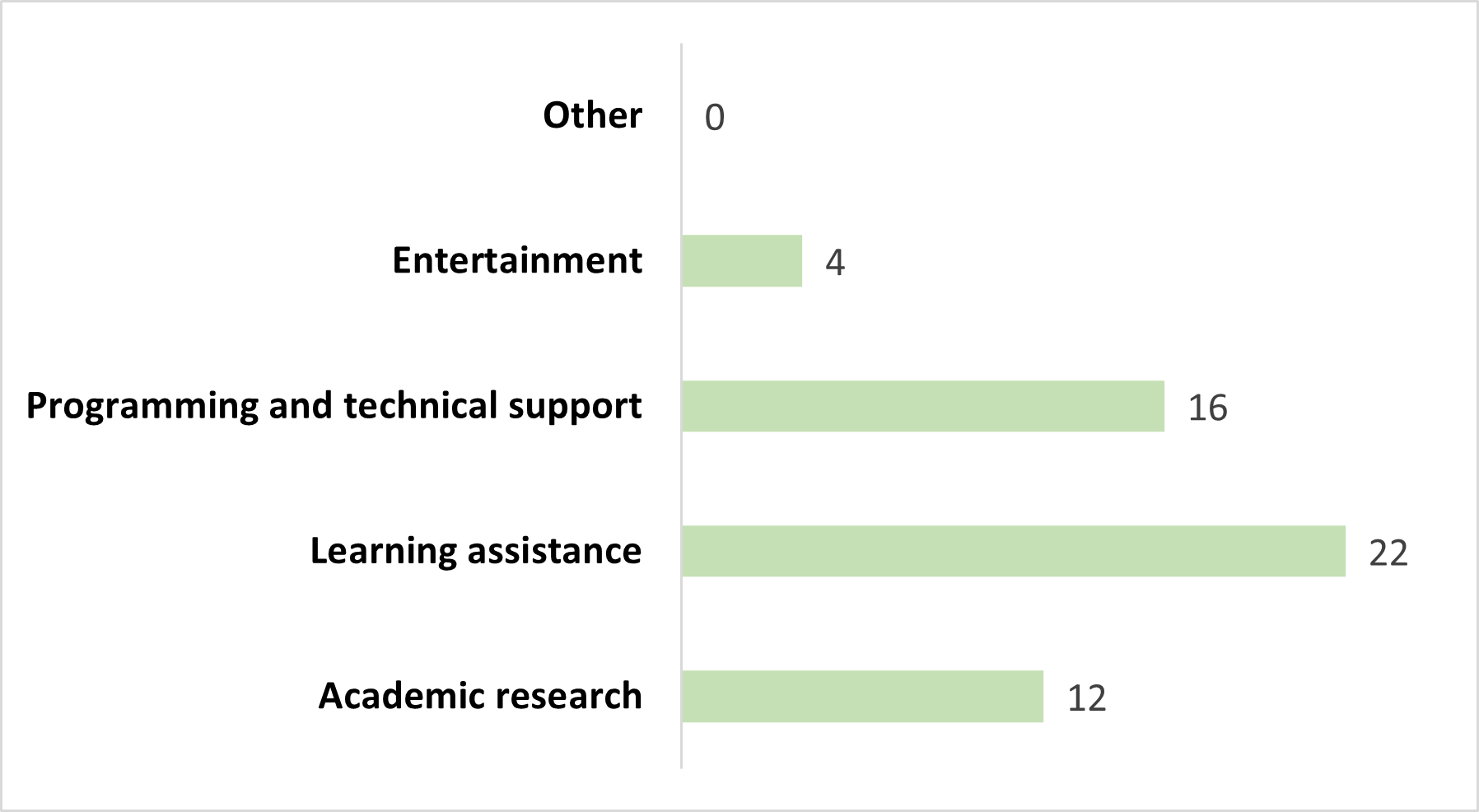} 
  \caption{\centering Results for Question (2): \textit{"In which scenarios do you primarily use LLMs? (Multiple choice)"}}
  \label{fig6}
\end{figure*}

\begin{figure*}[htbp]
  \centering
  \includegraphics[width=0.5\textwidth]{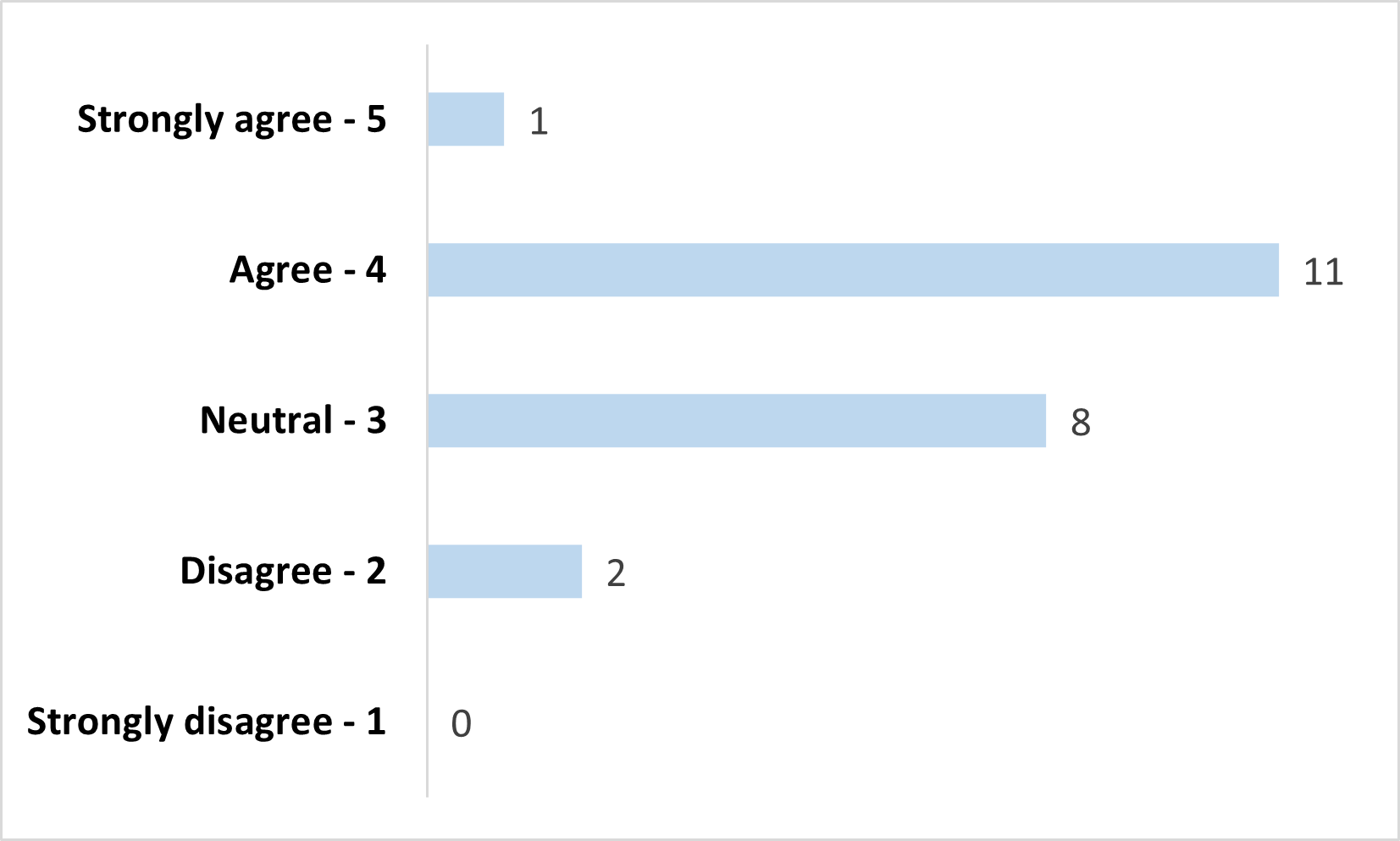} 
  \caption{\centering Results for Question (3): \textit{"Do you have experience in using LLMs and are capable of effectively using them to accomplish tasks?"}}
  \label{fig7}
\end{figure*}

The survey questionnaire was completed by 22 participants. The results are presented in Figures \ref{fig5}, \ref{fig6}, and \ref{fig7}.

\section{Statistics for User Messages and GPT Responses}\label{appendix_1}

The columns of Table \ref{tab:debate_statistics} are defined as follows: \textbf{User Msgs} represents the total number of messages sent by the user during each debate. \textbf{User Words} represents the total word count of all messages sent by the user within a specific debate, while \textbf{GPT Words} represents the total word count of all responses generated by GPT in the same debate. \textbf{Avg Words/User (SD)} represents the average number of words per user message, with the standard deviation (SD) provided. Similarly, \textbf{Avg Words/GPT (SD)} represents the average number of words per GPT response, with the corresponding standard deviation.

\begin{table*}[h!]
\centering
\caption{Statistics for User Messages and GPT Responses}
\label{tab:debate_statistics}
\begin{tabular}{@{}lcccccc@{}}
\toprule
\textbf{Debate} & \textbf{User Msgs} & \textbf{User Words} & \textbf{GPT Words} & \textbf{Avg Words/User (SD)} & \textbf{Avg Words/GPT (SD)} \\ \midrule
Debate 1        & 21                 & 2898                & 12209              & 138 (SD = 355)             & 581 (SD = 237)             \\
Debate 2        & 11                 & 3195                & 7691               & 290 (SD = 444)             & 699 (SD = 189)             \\
Debate 3        & 27                 & 2088                & 17852              & 77 (SD = 188)              & 661 (SD = 391)             \\ \bottomrule
\end{tabular}
\end{table*}
\end{document}